\documentclass[twocolumn]{aastex63}

\usepackage{amsmath}
\usepackage{colortbl}

\begin{document}

\title{Limits on hierarchical black hole mergers from the most negative $\chi_\mathrm{eff}$ systems}
\author{Maya Fishbach}
\altaffiliation{NASA Hubble Fellowship Program Einstein Fellow}
\affiliation{Center for Interdisciplinary Exploration and Research in Astrophysics (CIERA) and Department of Physics and Astronomy,
Northwestern University, 1800 Sherman Ave, Evanston, IL 60201, USA}
\author{Chase Kimball}
\affiliation{Center for Interdisciplinary Exploration and Research in Astrophysics (CIERA) and Department of Physics and Astronomy,
Northwestern University, 1800 Sherman Ave, Evanston, IL 60201, USA}
\author{Vicky Kalogera}
\affiliation{Center for Interdisciplinary Exploration and Research in Astrophysics (CIERA) and Department of Physics and Astronomy,
Northwestern University, 1800 Sherman Ave, Evanston, IL 60201, USA}

\begin{abstract}
     It has been proposed that some black holes (BHs) in binary black hole (BBH) systems are born from ``hierarchical mergers" (HM); i.e. earlier mergers of smaller BHs. These HM products have spin magnitudes $\chi \sim 0.7$, and, if they are dynamically assembled into BBH systems, their spin orientations will be sometimes anti-aligned with the binary orbital angular momentum. 
     In fact, as~\citet{2020PhRvD.102d3002B} showed, $\sim16\%$ of BBH systems that include HM products will have an effective inspiral spin parameter, $\chi_\mathrm{eff} < -0.3$. Nevertheless, the LIGO-Virgo-Kagra (LVK) gravitational-wave (GW) detectors have yet to observe a BBH system with $\chi_\mathrm{eff} \lesssim -0.2$, leading
     to upper limits on the fraction of HM products in the population.
     We fit the astrophysical mass and spin distribution of BBH systems and measure the fraction of BBH systems with $\chi_\mathrm{eff} < -0.3$, which implies an upper limit on the HM fraction. 
     We find that fewer than $26\%$ of systems in the underlying BBH population include HM products (90\%. credibility). Even among BBH systems with primary masses $m_1=60\,M_\odot$, the HM fraction is less than 69\%, which may constrain the location of the pair-instability mass gap.
     With 300 GW events (to be expected in the LVK's next observing run), if we fail to observe a BBH with $\chi_\mathrm{eff} < -0.3$, we can conclude that the HM fraction is smaller than $2.5^{+9.1}_{-2.2}\%$.
\end{abstract}

\section{Introduction}
\label{sec:intro}

The network of gravitational-wave detectors Advanced LIGO~\citep{2015CQGra..32g4001L}, Advanced Virgo~\citep{2015CQGra..32b4001A} and Kagra~\citep{2021PTEP.2021eA101A} has revealed a new population of binary black holes (BBHs), with 69 confident BBH events (false alarm rate $< 1$/year) in the latest LIGO-Virgo-Kagra (LVK) catalog~\citep{2021arXiv211103606T}. Although the BBH discovery rate is accelerating, the formation history of these BBH systems remains unknown. Broadly, a BBH system may either evolve from a pair of binary stars (``isolated binary evolution";~\citealp{1998ApJ...506..780B,2002ApJ...572..407B,2007PhR...442...75K,2012ApJ...759...52D}), or it may be assembled dynamically in a dense stellar environment containing many BHs (``dynamical assembly";~\citealp{1993Natur.364..421K,1993Natur.364..423S,2000ApJ...528L..17P}); see~\citet{2021hgwa.bookE...4M} and~\citet{2022PhR...955....1M} for recent reviews.

A popular way of distinguishing BBH formation channels is based on the statistics of their spin orientations~\citep{2016ApJ...832L...2R,2017CQGra..34cLT01V,2017PhRvD..96b3012T,2017MNRAS.471.2801S}. To first order, dynamical assembly leads to random spin orientations (isotropic tilts;~\citealp{2000ApJ...528L..17P}) while isolated binary evolution leads to preferential alignment between the BH spins and the orbital angular momentum axis~\citep{2000ApJ...541..319K}.
The spin magnitudes and orientations of the component BHs in a BBH affect the GW signal primarily through a combination known as the effective inspiral spin parameter, which is approximately conserved during the GW inspiral~\citep{2008PhRvD..78d4021R,2010PhRvD..82f4016S,2011PhRvL.106x1101A},
\begin{equation}
    \chi_\mathrm{eff} = \frac{m_1\chi_{1,z} + m_2\chi_{2,z}}{m_1 + m_2}.
\end{equation}
Here, $\chi_{1,z}$ ($\chi_{2,z}$) is the primary (secondary) dimensionless spin vector projected along the orbital angular momentum axis. 
The distribution of $\chi_\mathrm{eff}$ in the BBH population provides a simple test of BBH formation channels.
If spin tilts are isotropically distributed in BBH systems, the $\chi_\mathrm{eff}$ distribution will be symmetric about zero, pointing to a dynamical origin, whereas if spins are always aligned within $90^\circ$, the $\chi_\mathrm{eff}$ distribution will only have positive support, pointing to an isolated origin~\citep{2017Natur.548..426F}. 

A population analysis of the latest GW observations suggests that the BBH $\chi_\mathrm{eff}$ distribution is not symmetric about zero, with the majority of systems having $\chi_\mathrm{eff} > 0$~\citep{2021arXiv211103634T}. The observation of individual events with strictly positive $\chi_\mathrm{eff} > 0.2$ requires the $\chi_\mathrm{eff}$ distribution to extend to nonzero, positive $\chi_\mathrm{eff}$. In particular, the asymmetry in the $\chi_\mathrm{eff}$ distribution implies that the spin tilt distribution prefers tilts smaller than $90^\circ$, indicating that dynamical assembly cannot account for all BBH systems. {This asymmetry also suggests that some BHs in the isolated channel have nonzero spins, although the mechanism by which they acquire such spins remains poorly understood, depending on theoretically uncertain factors including angular momentum transport in massive stars~\citep{1973MNRAS.165...39T,2002A&A...381..923S,2019ApJ...870L..18Q,2019MNRAS.485.3661F}, tidal spin up of stripped stars in binary systems~\citep{2016MNRAS.462..844K,2018A&A...616A..28Q,2020A&A...635A..97B,2022MNRAS.511.3951F}, BH spin up during a supernova~\citep{2017ApJ...846L..15B,2018ApJ...862L...3S,2022ApJ...926....9J}, and BH spin up by accretion from its binary companion~\citep{2002ApJ...565.1107P,2015ApJ...800...17F,2008ApJ...689L...9M,2020ApJ...897..100V,2022ApJ...933...86Z}.}

Meanwhile, it remains unclear whether the $\chi_\mathrm{eff}$ distribution has support at negative $\chi_\mathrm{eff}$, which would suggest \emph{some} contribution from dynamical assembly.
The latest LVK catalog GWTC-3 does not contain any high-significance events with confidently negative $\chi_\mathrm{eff}$ (at $> 90\%$ credibility under the default parameter estimation priors).
However, a few GWTC-3 events, GW191109\_010717 and GW200225\_060421 have $\chi_\mathrm{eff} < 0$ with 90\% and 85\% credibility, respectively~\citep{2021arXiv211103606T}, and lower-significance candidate events with negative $\chi_\mathrm{eff}$ have been reported in independent analyses~\citep{2020PhRvD.101h3030V,2022arXiv220102252O}. 
Combining information across multiple events, \citet{2021ApJ...913L...7A,2021arXiv211103634T} and \citet{2022arXiv220508574C} find that the BBH population likely contains systems with (small) negative $\chi_\mathrm{eff}$ ($\chi_\mathrm{eff} \lesssim -0.02$), but this evidence weakens under population models that include a correlation between $\chi_\mathrm{eff}$ and mass ratio~\citep{2021arXiv210600521C}, or a nonspinning subpopulation that would create a narrow peak at zero in the $\chi_\mathrm{eff}$ distribution~\citep{2021arXiv210510580R,2021arXiv210902424G}. Disentangling small negative $\chi_\mathrm{eff}$ from nonspinning systems requires resolving the $\chi_\mathrm{eff}$ distribution to very small scales $\mathcal{O}(10^{-2})$~\citep{2022arXiv220508574C}. If dynamically-assembled BBHs always consist of slowly-spinning, or even nonspinning BHs~\citep{2019ApJ...881L...1F}, the ambiguity around the presence of systems with negative $\chi_\mathrm{eff}$ will persist. However, there is probably a subpopulation of spinning BHs in dense stellar environments: BHs born from previous mergers. 

In dynamical formation, as long as the escape speed of the environment is larger than typical GW recoil kicks, some BBH merger products will be retained and dynamically assembled into new BBH systems in so-called hierarchical mergers (HMs)~\citep{2016ApJ...831..187A,2016MNRAS.459.3432M,2018ApJ...866...66M,2018PhRvD..98l3005R,2022ApJ...927..231F}; see \citet{2021NatAs...5..749G} for a review. For example, HMs are typically predicted to account for $\sim10\%$ of mergers from globular clusters~\citep{2019PhRvD.100d3027R}, and this fraction increases for higher-density environments like nuclear star clusters~\citep{2021MNRAS.505..339M}. 

Furthermore, HMs are a promising explanation for some of the most massive BBHs observed by the LVK~\citep{2020ApJ...900L..13A,2021ApJ...915L..35K}, especially those with masses above $\sim40$-$65\,M_\odot$, where we expect few, if any, BHs born directly from stellar collapse due to the theorized pair-instability supernova mass gap~\citep{1964ApJS....9..201F,1984ApJ...280..825B,2002ApJ...567..532H,2016A&A...594A..97B}. GW observations find an excess of BH component masses between $\sim35$--$40\,M_\odot$, followed by a steeper decrease in the merger rate, which may be a signature of pair instability~\citep{2017ApJ...851L..25F,2018ApJ...856..173T,2021ApJ...913L...7A,2021arXiv211103634T}. BBHs with primary masses above this excess may then be HM products. The merger rate of BBHs with primary masses between $50$--$100\,M_\odot$ is $0.099$--$0.4\,\mathrm{Gpc}^{-3}\,\mathrm{yr}^{-1}$ ($\lesssim1\%$ of the total merger rate, \citealp{2021arXiv211103634T}).

In the HM scenario, the binary's orbital angular momentum contributes to the spin of the merger product, thereby producing BBHs with spinning components and, under the assumption of a classical star cluster\footnote{We focus on classical star clusters in this work, but gas-rich stellar environments, such as the disks of active galactic nuceli (AGN), may produce preferentially aligned or misaligned spin tilts due to gas torques, as we discuss in the following sections~\citep{2007ApJ...661L.147B,2018ApJ...866...66M}.\label{foot:agn}}, isotropic spin tilts~\citep{2017ApJ...840L..24F,2017PhRvD..95l4046G,2020ApJ...893...35D}. BBH systems that contain a HM product therefore have a broad distribution of $\chi_\mathrm{eff}$ that is symmetric about zero~\citep{2020RNAAS...4....2K,2020PhRvD.102d3002B}. In particular, some fraction of these BBH systems must exhibit significantly negative $\chi_\mathrm{eff}$.

In this paper, {we find that there is a threshold of $\chi_\mathrm{eff} < -0.3$, below which we always expect to find the same fraction  ($\sim16\%$) of HMs, regardless of the initial BH spins, binary mass ratio, or number of previous mergers. We then use LVK observations to place upper limits on the fraction of BBH systems with $\chi_\mathrm{eff} < -0.3$ in the underlying astrophysical population. Comparing this inferred upper limit against the expected fraction of HMs with $\chi_\mathrm{eff} < -0.3$ constrains the HM contribution to the BBH population.}
As~\citet{2020PhRvD.102d3002B} showed, if repeated mergers are the only way for BBHs to gain large spins, this upper limit {(and a corresponding limit on observations with large, positive $\chi_\mathrm{eff}$)} becomes a direct measurement of the fraction of HMs in the population. {However, as discussed earlier, both theory and observation suggest that BBHs can acquire spins in alternative scenarios, especially through binary interactions in the isolated formation channel, which may explain BBH observations with $\chi_\mathrm{eff} > 0.2$. Among these spin-up scenarios, HMs are unique in that they also invariably produce large, negative $\chi_\mathrm{eff}$. Thus, the absence of events with $\chi_\mathrm{eff} < -0.3$ constrains the HM contribution more tightly than using the positive $\chi_\mathrm{eff}$ fraction, yet robustly enough to remain applicable across a wide range of HM scenarios.}

The remainder of this paper is organized as follows. Section~\ref{sec:chieff-dists} derives theoretical $\chi_\mathrm{eff}$ distributions for BBH systems that contain one or two HM products, showing that we generically expect 16\% of these systems to have $\chi_\mathrm{eff} < -0.3$. In Section~\ref{sec:results}, we infer the astrophysical $\chi_\mathrm{eff}$ distribution from LVK observations and place constraints on the HM fraction, showing that even among BBHs with primary masses of $60\,M_\odot$ (possibly within the pair-instability mass gap), HMs make up fewer than 69\% of systems. We discuss expectations for future observations in Section~\ref{sec:future} and conclude in Section~\ref{sec:conclusion}.

\section{Effective inspiral spins from hierarchical mergers}
\label{sec:chieff-dists}

\begin{figure}
    \centering
    \includegraphics[width = 0.5\textwidth]{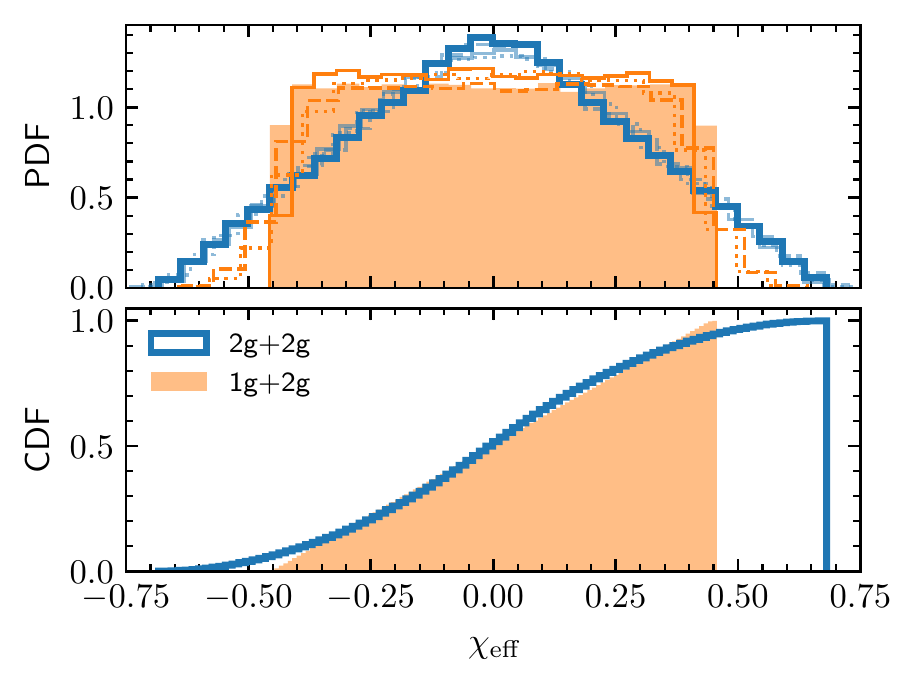}
    \caption{Theoretical distribution of $\chi_\mathrm{eff}$ among BBHs in which one (orange) or both (blue) components are the products of previous mergers. \emph{Top panel:} $\chi_\mathrm{eff}$ probability distribution function for four different assumptions about the initial spins $\chi_\mathrm{1g}$ and mass ratios $q_\mathrm{1g}$ of 1g+1g black holes: $\chi_\mathrm{1g} = 0$ and a uniform $q_\mathrm{1g}$ distribution $0.8 < q_\mathrm{1g} < 1$  (default; thick blue and filled orange); $\chi_\mathrm{1g} = 0$ and $0.5 < q_\mathrm{1g} < 1$ (thin solid lines); uniform $0 < \chi_\mathrm{1g} < 0.5$ and $0.8 < q_\mathrm{1g} < 1$ (thin dashed lines); $0 < \chi_\mathrm{1g} < 0.5$ and $0.5 < q_\mathrm{1g} < 1$ (thin dotted lines). The mass ratios of mixed-generation 1g+2g mergers are always assumed to be given by ${1}/{(1 + q_{1\mathrm{g}})}$ where $q_{1\mathrm{g}}$ is drawn from the assumed mass ratio distribution for 1g+1g mergers. \emph{Bottom panel:} Corresponding cumulative distribution functions, shown just for the default spin and mass ratio assumptions.}
    \label{fig:chieff_PDF_CDF_HM}
\end{figure}

A merger of two first-generation (1g+1g) BHs of comparable masses (mass ratio $0.5 < q_{1g} < 1$) produces a second-generation (2g) BH with dimensionless spin magnitude $\chi \sim 0.7$~\citep{2008ApJ...684..822B,2008PhRvD..77b6004B,2008PhRvD..78h1501T,2017ApJ...840L..24F}. 
If this 2g BH dynamically forms a binary and merges with another similarly-formed 2g BH, the effective inspiral spin of this 2g+2g BBH merger (with mass ratio $q_{2g} = q_{1g}$) will be drawn from the probability distribution function (PDF) shown in the blue, unfilled histogram in the top panel of Fig.~\ref{fig:chieff_PDF_CDF_HM}, with corresponding cumulative distribution function (CDF) shown in the bottom panel. Meanwhile, if the 2g BH instead merges with one of the original 1g BHs, the effective inspiral spin of the resulting 1g+2g BBH (with mass ratio $1/(1+q_{1g}$) will be drawn from the PDF shown by the orange, filled histogram in Fig.~\ref{fig:chieff_PDF_CDF_HM}. The $\chi_\mathrm{eff}$ distributions of 2g+2g and 1g+2g BBHs were previously derived in~\citet{2020PhRvD.102d3002B}. Here, we build the $\chi_\mathrm{eff}$ distribution with Monte Carlo samples. We first draw 1g+1g BBHs according to three different mass ratio $q_{1g}$ and spin $\chi_{1g}$ distributions, as explained in the caption of Fig.~\ref{fig:chieff_PDF_CDF_HM}. {These different 1g+1g distributions have mass ratios in the range $0.5 < q_{1g} < 1$ and spins in the range $0 < \chi_{1g} < 0.5$, motivated by both theoretical expectations and BBH observations that suggest asymmetric mass ratios and high spins are rare among merging systems~\citep{2021ApJ...913L...7A,2021arXiv211103634T,2022PhR...955....1M}.} We then calculate 2g BH spins with \textsc{PESummary}~\citep{Hoy:2020vys}, which averages over numerical relativity fits~\citep{2014PhRvD..90j4004H,2016ApJ...825L..19H,2017PhRvD..95f4024J}. We assume BBHs always merge with isotropic spin orientations.

Figure~\ref{fig:chieff_PDF_CDF_HM} shows that for both 2g+2g and 1g+2g mergers, $\chi_\mathrm{eff}$ follows a broad distribution, with significant support at $|\chi_\mathrm{eff}| > 0.3$. The $\chi_\mathrm{eff}$ distribution depends only slightly on the mass ratios $q_{1g}$ and initial spins of 1g+1g BHs. {The near-universality of the HM $\chi_\mathrm{eff}$ distribution stems from the fact that merger products always have $\chi_\mathrm{2g} \sim 0.7$, for a wide range of initial spins and mass ratios~\citep{2017ApJ...840L..24F}.} Our assumption of isotropic spin tilts implies that the $\chi_\mathrm{eff}$ distribution is always symmetric about zero. In both 2g+2g and 1g+2g scenarios, approximately $16\%$ of systems assembled hierarchically have $\chi_\mathrm{eff} < -0.3$ (ranging from $15$--$17\%$ for the scenarios shown in Fig.~\ref{fig:chieff_PDF_CDF_HM}). %For 2g+2g mergers, which include two HM products, $|\chi_\mathrm{eff}|$ can extend to larger values, and $8$--$9$\% of 2g+2g BBH mergers have $\chi_\mathrm{eff} < -0.4$.
This result agrees with~\citet{KimballInPrep}, who update the results of~\citet{2021ApJ...915L..35K} with GWTC-3 and fit the mass and spin distribution for first-generation and HM BBH systems, together with the branching fraction between the two sub-populations. They find that $15.4\%$ of BBHs among their HM sub-population have $\chi_\mathrm{eff} < -0.3$.

Higher generations of BHs will follow similar distributions because the final spin stays close to $\sim0.7$, {even if the individual merging BHs have large spins}. The exception is BH growth through repeated minor mergers ($q \ll 1$), which tend to spin the BH down~\citep{2003ApJ...585L.101H}. As mentioned in footnote~\ref{foot:agn}, our analysis also does not strictly apply to gas-rich environments for dynamical assembly, such as AGN disks, where the distribution of spin tilts may not be isotropic, although misaligned and anti-aligned systems are expected to be common~\citep{2007ApJ...661L.147B,2018ApJ...866...66M,2020ApJ...899...26T,2022ApJ...931...82V}. In this case, if BHs grow through many generations of preferentially-aligned (anti-aligned), repeated mergers, the spin magnitude of HM products will converge to $\chi \sim 0.9$ ($\chi \sim 0.5$) rather than $\chi \sim 0.7$~\citep{2017ApJ...840L..24F}. Additionally, the $\chi_\mathrm{eff}$ distribution may not be symmetric. Nevertheless, we expect HMs in AGN disks to produce BBH systems with large, negative $\chi_\mathrm{eff}$~\citep{2020ApJ...899...26T}, and so our method broadly applies to this scenario, although the quantitative details depend on the highly uncertain spin tilt distribution of BBH systems in AGN disks.

We can convert the inferred fraction of systems with $\chi_\mathrm{eff} < -0.3$, $f_{\chi_\mathrm{eff} < -0.3}$, into an upper limit on the fraction of HM systems $f_\mathrm{HM}$ (including both 2g+2g and 1g+2g systems) in the astrophysical BBH population according to the simple rule,
\begin{equation}
\label{eq:fHM}
    f_\mathrm{HM} \leq \frac{1}{0.16}f_{\chi_\mathrm{eff} < -0.3} = 6.25 f_{\chi_\mathrm{eff} < -0.3}.
\end{equation}
Meanwhile, the fraction of systems with $\chi_\mathrm{eff} < 0$ can be used to infer an upper limit on the fraction of dynamically assembled BBHs $f_\mathrm{dyn}$,
\begin{equation}
\label{eq:fdyn}
    f_\mathrm{dyn} \leq \frac{1}{0.5}f_{\chi_\mathrm{eff} < 0} = 2 f_{\chi_\mathrm{eff} < 0}.
\end{equation}

\section{Observational limits on the negative $\chi_\mathrm{eff}$ fraction}
\label{sec:results}

\begin{figure*}
    \centering
    \includegraphics[width = 0.95\textwidth]{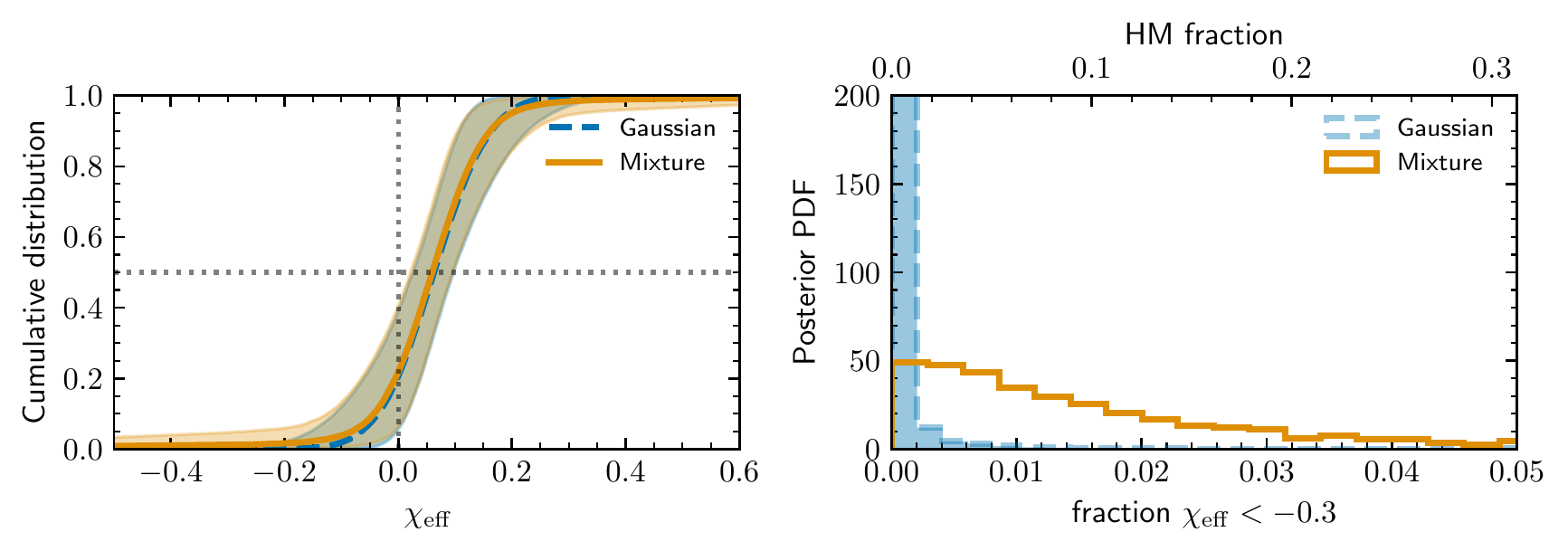}
    \caption{Inferred $\chi_\mathrm{eff}$ CDF for the astrophysical BBH population (left) and posterior PDF on the fraction of $\chi_\mathrm{eff} < -0.3$ (right). We infer the $\chi_\mathrm{eff}$ distribution by fitting the confident BBH events in GWTC-3 to two models, a Gaussian (blue) and a mixture model between a Gaussian and a uniform distribution (orange), both truncated to the physically allowed range $-1 < \chi_\mathrm{eff} < 1$. Solid lines show the median CDF value at each $\chi_\mathrm{eff}$ and shaded bands show 90\% credible intervals. Both models assume that the $\chi_\mathrm{eff}$ distribution does not vary with BBH mass. The mixture model is designed to allow for more support at the tails of the $\chi_\mathrm{eff}$ distribution, and so allows for a higher fraction of systems with $\chi_\mathrm{eff} < -0.3$. The constraints on the $\chi_\mathrm{eff} < -0.3$ fraction can be interpreted as limits on the contribution of HMs to the astrophysical BBH population (right panel, top axis).}
    \label{fig:chieff_GaussT}
\end{figure*}

The $\chi_\mathrm{eff}$ CDF for the astrophysical BBH population, as inferred from the 69 confident BBH events in GWTC-3, is shown in the right panel of Fig.~\ref{fig:chieff_GaussT}. In blue, we show the $\chi_\mathrm{eff}$ distribution inferred under a truncated Gaussian model~\citep{2019MNRAS.484.4216R,2020ApJ...895..128M}, with flat priors on the mean (between -0.5 and 0.5) and standard deviation (between 0.03 and 0.5). We assume that the $\chi_\mathrm{eff}$ distribution is independent of other BBH properties like mass or redshift, and we simultaneously fit the primary mass, mass ratio, and redshift distributions. We adopt a broken power law model for the primary mass distribution, a power law model for the mass ratio conditioned on the primary mass, and a power law in $(1+z)$ for the redshift distribution~\citep{2021ApJ...913L...7A}. While we expect the mass ratio distribution to have some effect on the $\chi_\mathrm{eff}$ distribution because of the correlations between $\chi_\mathrm{eff}$ and mass ratio in the parameter estimation of individual events, we do not expect the details of the primary mass or redshift distribution to noticeably affect our results. To perform the fit, we use the parameter estimation samples released in~\citet{2019PhRvX...9c1040A, 2021PhRvX..11b1053A,2021arXiv210801045T,2021arXiv211103606T}, using the available ``Overall\_posterior" (for the first two observing runs), ``NRSur7dq4", ``PrecessingSpinIMRHM" or ``C01:Mixed" (for the third observing run) set of samples. We use the detector sensitivity estimates from~\citet{2021arXiv211103606T,ligo_scientific_collaboration_and_virgo_2021_5636816}. 

The inferred distribution is consistent with the results of~\citet{2021arXiv211103634T}, who fit the truncated Gaussian model to the same set of events. The median $\chi_\mathrm{eff}$ (a CDF value of 0.5, denoted by the dashed horizontal line) is greater than zero (denoted by the dashed vertical line) at $>99\%$ credibility, implying that the $\chi_\mathrm{eff}$ distribution is not symmetric about zero, and not all binaries are dynamically assembled. In fact, following Eq.~\ref{eq:fdyn}, the inferred $\chi_\mathrm{eff} < 0$ fraction implies a dynamically assembled fraction of at most $0.42^{+0.37}_{-0.31}$ (90\% credibility). 

Under the Gaussian model, we infer $f_{\chi_\mathrm{eff} < -0.3} < 0.08\%$ (see right panel of Fig.~\ref{fig:chieff_GaussT}), which,  according to Eq.~\ref{eq:fHM}, suggests that HMs comprise fewer than 0.5\% of the BBH population (90\% credibility). Given that not all BBH systems are dynamically assembled, this upper limit corresponds to $f_\mathrm{HM}/f_\mathrm{dyn} = 0.7\%$ of the dynamically assembled population. However, this tight upper limit falsely assumes that we can resolve the tails of the $\chi_\mathrm{eff}$ distribution to sub-percent accuracy. In reality, because we only fit 69 events, we can only probe the distribution to a resolution of $1/69 \sim 1.4\%$, and conclude that fewer than 1.4\% of BBH events in the \emph{detected} population have $\chi_\mathrm{eff} \lesssim -0.2$. (According to the inferred $\chi_\mathrm{eff}$ distribution, the minimum observed $\chi_\mathrm{eff}$ out of 69 events is $\chi_\mathrm{eff} \gtrsim -0.2$). Among \emph{detected} HM events, assuming the HM fraction is the same at all BBH masses (an assumption that we revisit later), 11\% have $\chi_\mathrm{eff} < -0.3$. Thus, we expect a more conservative upper limit on the HM fraction would be $0.014/0.11 \sim \mathcal{O}(10\%)$, which is much higher than the $0.5\%$ inferred under the Gaussian model.

To derive a conservative upper limit on the HM fraction, we fit the $\chi_\mathrm{eff}$ distribution to a model inspired by~\citet{2021PhRvD.104h3010R}: a mixture model between a Gaussian and uniform component truncated to the range $-1 < \chi_\mathrm{eff} < 1$,
\begin{equation}
    p(\chi_\mathrm{eff} \mid \mu, \sigma, f_U) = \frac{f_U}{2} + (1 - f_U) \mathcal{N}^{T_{[-1, 1]}}_{\mu, \sigma}(\chi_\mathrm{eff}).
\end{equation}
We refer to this model as a Gaussian + Uniform Mixture. We take the same priors on $\mu, \sigma$ as the Gaussian model, and a flat prior on the mixing fraction $f_U$ in the range $0 < f_U < 1$. This model is introduced by~\citet{2021PhRvD.104h3010R} to allow for the presence of population outliers (which may or may not have already been observed) with respect to the ``bulk" population. In our case, the bulk population, which contains most of the observations, is the Gaussian component. The broad uniform component can accommodate any outlier BBH systems; in particular, HM systems. 
The Gaussian + Uniform Mixture model therefore favors distributions with support at extreme values of $\chi_\mathrm{eff}$ far from the mean of the Gaussian. In other words, the induced prior on $f_{\chi_\mathrm{eff} < -0.3}$ from the mixture model favors large values.

The $\chi_\mathrm{eff}$ CDF inferred under the Gaussian + Uniform Mixture model is shown in orange in Fig.~\ref{fig:chieff_GaussT} (left panel). The bulk 10--90\% of the inferred distribution agrees with results under a Gaussian model, but as expected, the Gaussian + Uniform Mixture model puts more support at the tails of the distribution. As shown in the right panel of Fig.~\ref{fig:chieff_GaussT}, under the mixture model, we measure that up to 3.7\% of BBH systems have $\chi_\mathrm{eff} < -0.3$ (90\% credibility). This implies an upper limit $f_\mathrm{HM} \lesssim 23\%$ out of the total BBH population, or 63\% of the dynamically assembled population, which, by design, is much more conservative than the result from the Gaussian model.

The results in Fig.~\ref{fig:chieff_GaussT} assume that the $\chi_\mathrm{eff}$ distribution does not depend on BBH mass. However, we expect the HM contribution to be larger among more massive BBH systems, causing a signature trend between $\chi_\mathrm{eff}$ and BBH mass~\citep{2020ApJ...894..129S,2020PhRvD.102d3002B,2021MNRAS.507.3362T,2022arXiv220113098F}. With GWTC-3, there is no conclusive evidence that the $\chi_\mathrm{eff}$ distribution varies with mass~\citep{2021arXiv211103634T,2022arXiv220401578B}, although it is not ruled out. To explore this possibility, we fit mass-dependent $\chi_\mathrm{eff}$ distributions and measure $f_{\chi_\mathrm{eff} < -0.3}(m_1)$ and the corresponding upper limit on $f_\mathrm{HM}(m_1)$ as functions of the primary BH mass $m_1$.
We extend both the Gaussian and Gaussian + Uniform Mixture models introduced previously to allow for a possible dependence on $m_1$. In the Gaussian model, we take the mean and standard deviations to be functions of $m_1$. In the Gaussian + Uniform Mixture model, we take the mixture fraction $f_U$ to be a function of $m_1$. We assume the same models for the marginal primary mass, mass ratio and redshift distributions as before.

In detail, in the $m_1$-dependent Gaussian model, we assume $\mu(m_1)$ and $\log\sigma(m_1)$ are linear. We define $\mu_{85} \equiv \mu(m_1 = 85\,M_\odot)$ and $\mu_{5} \equiv \mu(m_1 = 5\,M_\odot)$, and take
\begin{equation}
    \mu(m_1) = \frac{\mu_{85} - \mu_5}{80}(m_1 - 5) + \mu_5.
\end{equation}
Similarly, defining $\log\sigma_{85} \equiv \log\sigma(m_1 = 85\,M_\odot)$, and $\log\sigma_{5} \equiv \log\sigma(m_1 = 5\,M_\odot)$, we take
\begin{equation}
    \log\sigma(m_1) = \frac{\log\sigma_{85} - \log\sigma_5}{80} (m_1 - 5) + \log\sigma_5.
\end{equation}

In the $m_1$-dependent Gaussian + Uniform Mixture model, we define $f_5 \equiv f_U(m_1 = 5\,M_\odot)$ and $f_{85} \equiv f_U(m_1 = 85\,M_\odot)$, and take
\begin{equation}
    f_U(m_1) = \left(1 + a e^{bm_1}\right)^{-1},
\end{equation}
where
\begin{equation}
    b = 1/80 \left[{\log \left( 1/f_{85} - 1 \right) - \log \left( 1/f_5 - 1 \right)}\right]
\end{equation}
and
\begin{equation}
    a = \left( 1/f_5 - 1 \right)e^{-5b}.
\end{equation}
We take flat priors on all parameters.

\begin{figure}
    \centering
    \includegraphics[width=0.5\textwidth]{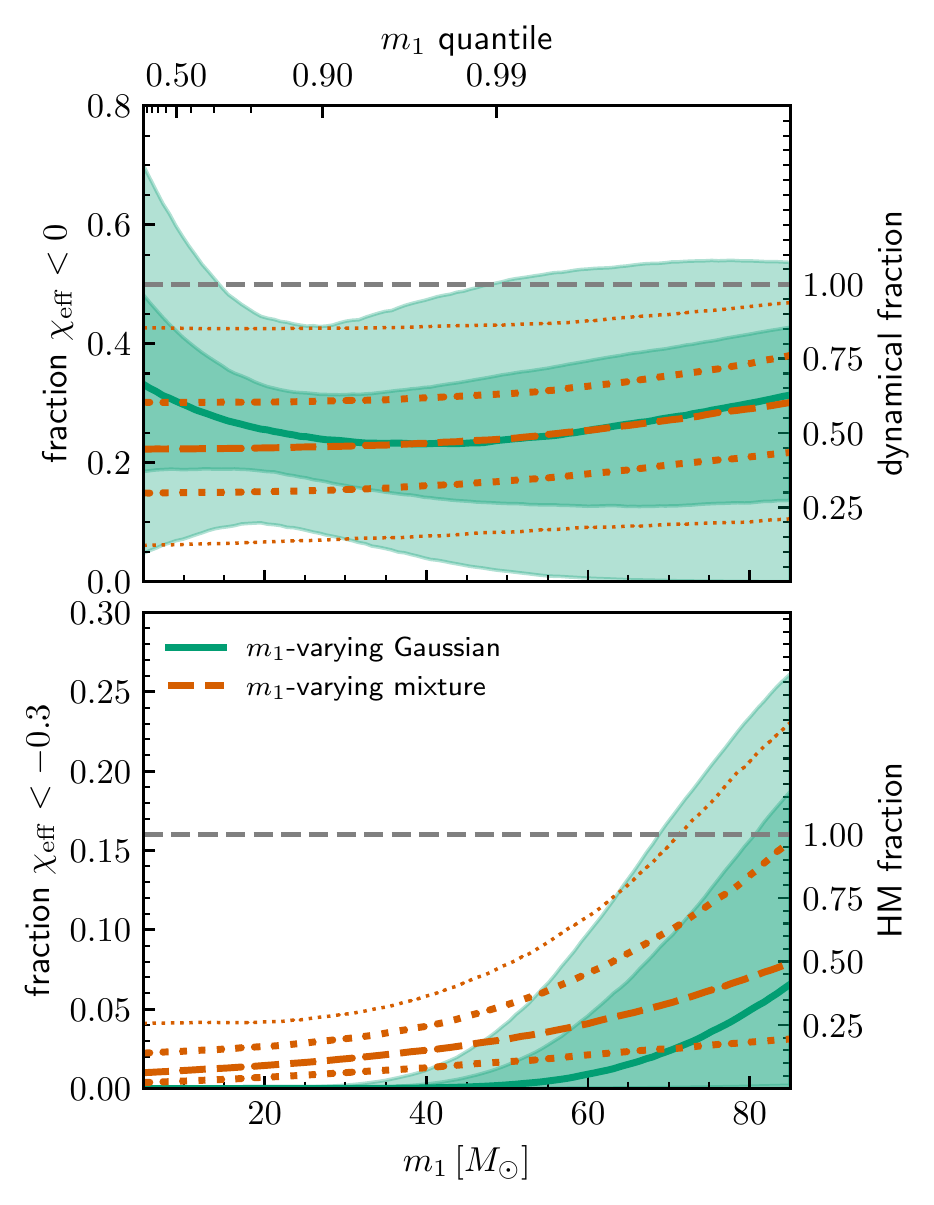}
    \caption{Fraction of systems with $\chi_\mathrm{eff} < 0$ (top panel) and $\chi_\mathrm{eff} < -0.3$ (bottom panel) at each primary mass $m_1$, inferred by fitting a mass-dependent $\chi_\mathrm{eff}$ distribution. The bottom axis shows primary mass in solar masses, while the top axis shows the corresponding quantile in the marginal $m_1$ posterior population distribution. In each panel, the orange dashed line shows the median value at each $m_1$ inferred under the $m_1$-varying Gaussian + Uniform mixture model, while the green solid line shows the median for the $m_1$-varying Gaussian model. Inner orange dotted lines (dark green shaded region) enclose the symmetric 50\% credible interval for the mixture (Gaussian) model. In the top panel, outer orange dotted lines (light green shaded region) enclose the symmetric 90\% credible interval for the two models, respectively, while in the bottom panel, we show the 90\% upper limit, because the posterior peaks at HM fraction = 0 for all $m_1$. Marginalizing over the mass distribution, the total HM fraction across all masses is less than 1.3\% (Gaussian model) or 26\% (mixture model; 90\% credibility).}
    \label{fig:negchieff-linevolm1}
\end{figure}

The resulting constraints on $f_{\chi_\mathrm{eff} < 0}(m_1)$, which places an upper limit on the dynamical fraction as a function of $m_1$, and $f_{\chi_\mathrm{eff} < -0.3}(m_1)$, which limits the HM fraction as a function of $m_1$, are shown in the top and bottom panels of Fig.~\ref{fig:negchieff-linevolm1}. 
As seen in the top panel, the dynamical fraction must be less than 1, at least among systems with primary masses in the range {$\sim10$--$45\,M_\odot$}, according to both the $m_1$-varying Gaussian and $m_1$-varying Gaussian + Uniform Mixture models. The $m_1$-varying Gaussian permits the dynamical fraction to extend to 1 at lower masses and higher masses, where there are fewer events and the constraints are weaker.

The bottom panel of Fig.~\ref{fig:negchieff-linevolm1} shows that the $f_{\chi_\mathrm{eff} < -0.3}$ is consistent with zero at all $m_1$, in agreement with the results of Fig.~\ref{fig:chieff_GaussT}. Nevertheless, under the greater flexibility of the $m_1$-varying models, it is possible that $f_{\chi_\mathrm{eff} < -0.3}$, and therefore the HM contribution, increases with increasing primary mass, as we would expect in a HM scenario. The results are similar for both the $m_1$-varying models of Fig.~\ref{fig:negchieff-linevolm1}, although the mixture model \emph{a priori} favors larger HM fractions and provides a more conservative upper limit. Our conclusions are consistent with the mass-dependent $\chi_\mathrm{eff}$ fits of~\citet{2022arXiv220113098F} (see their Fig. 3). However, in agreement with~\citet{2022arXiv220401578B}, we do not find strong evidence that the $\chi_\mathrm{eff}$ distribution varies with mass. Instead, the data are consistent with a constant (small) HM fraction at all masses, but permit a larger HM fraction at higher masses where there are fewer events, leading to weaker upper limits. 

We find that among BBHs with primary masses $m_1 = 60\,M_\odot$, the HM fraction is less than 64\% (according to the $m_1$-varying Gaussian model) or 69\% (according to the $m_1$-varying Gaussian + Uniform mixture model), at 90\% credibility. The fact that this fraction is smaller than unity implies that, in the absence of a more exotic alternative formation channel, stellar collapse must make BHs with masses up to $\gtrsim 60\,M_\odot$. This limit on the location of the pair-instability mass gap in turn has implications for stellar structure~\citep{2020MNRAS.493.4333R,2021MNRAS.501.4514C,2021ApJ...912L..31W,2020ApJ...905L..15B}, nuclear physics~\citep{2019ApJ...887...53F,2020ApJ...902L..36F}, new particles~\citep{2020arXiv200707889C} and cosmology~\citep{2019ApJ...883L..42F}.

Although HMs may make up the majority of the high-mass population with $m_1 \gtrsim 60\,M_\odot$, systems at these masses make up less than $\sim1\%$ of the underlying population, as shown in the top axis of Fig.~\ref{fig:negchieff-linevolm1}. The small contribution of these high-mass systems, for which HMs may dominate, to the overall BBH population implies that HMs must make up no more than 26\% of the total BBH population (at 90\% credibility), similar to the upper limit of 23\% inferred under the $m_1$-independent model. If the HM fraction is as high as $26\%$, the majority of HMs are hidden below $50$--$60\,M_\odot$, because the high-mass BBH systems can only account for 1\% of mergers.

We have presented results on the HM fraction using a Gaussian model and a Gaussian + Uniform Mixture model, but we have also verified that our upper limit of 26\% holds with alternative models for the $\chi_\mathrm{eff}$ distribution. For example, we considered a Student T distribution (which has more support at the tails compared to a Gaussian), as well as mixture models between a Gaussian and one of the predicted HM distributions of Fig.~\ref{fig:chieff_PDF_CDF_HM}. In this latter case, the inferred branching fraction for the HM component can be directly interpreted as the HM fraction, and agrees with our upper limit of $f_\mathrm{HM} \lesssim 26\%$ within a couple of percent regardless of the assumed HM distribution. Furthermore, this limit is consistent, but more conservative than, the inferred upper limit on the HM fraction from~\citet{KimballInPrep}.

\section{Predictions for future observations}
\label{sec:future}

\begin{figure}
    \centering
    \includegraphics[width=0.5\textwidth]{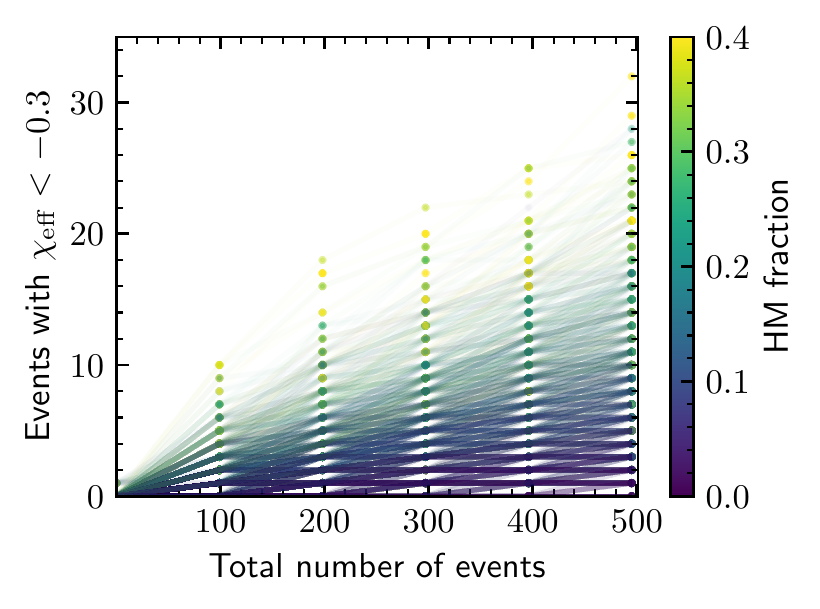}
    \caption{Expected number of BBH events with $\chi_\mathrm{eff}$ as the total number of observations increases. We plot 2000 realizations, where each realization is drawn from the hyperposterior of the $m_1$-varying Gaussian + Uniform Mixture model. Each realization is colored by the corresponding HM fraction. Larger HM fractions predict more events with $\chi_\mathrm{eff} < -0.3$, especially as the total number of events approaches 500.}
    \label{fig:chineg_obs_vtotal_HMfrac.pdf}
\end{figure}

With 69 BBH events, the absence of observations with $\chi_\mathrm{eff} < -0.3$ limits the contribution of HMs to be most likely under $\mathcal{O}(10\%)$, but HM fractions as high as $\sim 30\%$ are still permitted by the data. As the BBH catalog grows, the number of future observations with $\chi_\mathrm{eff} < -0.3$ will inform a tighter measurement of the HM contribution. Figure~\ref{fig:chineg_obs_vtotal_HMfrac.pdf} shows the expected number of events with $\chi_\mathrm{eff} < -0.3$ as a function of the total number of observations for 2000 different realizations, where each realization corresponds to a different hyperposterior draw under the $m_1$-varying Gaussian + Uniform Mixture model. A hyperposterior draw specifies a mass-spin-redshift distribution from which we in turn draw 500 mock observations. After every 100 such mock observations, we track how many of them have $\chi_\mathrm{eff} < -0.3$, as plotted in Fig.~\ref{fig:chineg_obs_vtotal_HMfrac.pdf}. Each line is colored by the total HM fraction of its corresponding hyperposterior draw; there are more realizations with small HM fractions (dark purple curves) because we inferred that these values are more likely. When drawing mock observations, we incorporate selection effects matching the LVK detector sensitivity across the first three observing runs~\citep{ligo_scientific_collaboration_and_virgo_2021_5636816}. This is an approximation for the detector sensitivity in future observing runs, which will observe events to much higher distances, but will have similar mass- and $\chi_\mathrm{eff}$-dependent selection effects. 

As seen in Fig.~\ref{fig:chineg_obs_vtotal_HMfrac.pdf}, large HM fractions generally predict a higher fraction of events with $\chi_\mathrm{eff} < -0.3$. We currently have fewer than 100 total events, and so we would not expect more than a couple of events with $\chi_\mathrm{eff} < -0.3$ according to any of the realizations (by construction, all of the allowed HM fractions are consistent with the current lack of events with $\chi_\mathrm{eff} < -0.3$). However, by the time we observe 300 events, if the HM fraction is higher than 20\%, we expect to observe $7^{+7}_{-5}$ events with a true $\chi_\mathrm{eff} < -0.3$; we will observe at least 1 such event with 99.4\% credibility. Meanwhile, if the HM fraction is smaller than 10\%, unless there is another mechanism for producing events with large negative $\chi_\mathrm{eff}$, we will observe fewer than 6 such events (90\% credibility). 

A related question is whether we will observe events for which we can confidently identify that $\chi_\mathrm{eff}$ is negative despite typical measurement uncertainties. A population analysis reveals the fraction of events with negative $\chi_\mathrm{eff}$, but not necessarily which individual events have $\chi_\mathrm{eff} < 0$, especially if $|\chi_\mathrm{eff}|$ is small. However, events for which the true $\chi_\mathrm{eff}$ is more negative than $-0.3$ will usually have very little likelihood support at positive $\chi_\mathrm{eff}$, allowing us to identify $\chi_\mathrm{eff} < 0$ at high ($>99\%$) credibility~\citep{2018PhRvD..98h3007N}. Figure~\ref{fig:chineg_obs_vtotal_HMfrac.pdf} suggests that if the HM fraction is above 10\%, we expect an observation with confidently negative $\chi_\mathrm{eff}$ within the first 150 observations. If we have not seen a system with a confidently negative $\chi_\mathrm{eff}$ within the first 300 events, we may conclude that the HM fraction is smaller than $2.5^{+9.1}_{-2.2}\%$.

\section{Conclusion}
\label{sec:conclusion}
The formation of BBH systems through dynamical interactions will generally lead to misaligned spins, with some systems having $\chi_\mathrm{eff} < 0$. Furthermore, because repeated mergers produce BHs with spin magnitudes $\chi \sim 0.7$, HMs in dense stellar environments will produce BBH systems with $\chi_\mathrm{eff} < -0.3$.

Among the current GW observations by the LVK Collaboration, some BBH events may have small negative $\chi_\mathrm{eff}$, but there are no observations with $\chi_\mathrm{eff} \lesssim -0.2$. We argue that the lack of observations with moderately large, negative $\chi_\mathrm{eff}$ values limits the possible contribution of second- or higher-generation HM systems to the BBH population. This in turn limits the escape speeds of dense stellar environments, because environments with high escape speeds will inevitably produce too many HMs to match the dearth of observations with $\chi_\mathrm{eff} \lesssim -0.2$. \citet{2022arXiv220508549Z} make a similar argument, using upper limits on the high-mass merger rate. They show that environments with high escape speeds will overproduce BBH mergers with masses above $\sim50\,M_\odot$ if BBH formation is too efficient. Here we use upper limits on the negative $\chi_\mathrm{eff}$ rate to place complementary limits on the HM fraction.

Our main conclusions are as follows:
\begin{itemize}
    \item As our most conservative upper limit, fewer than $4.2\%$ of BBH systems in the astrophysical population have $\chi_\mathrm{eff}  < -0.3$. Assuming a classical star cluster in which BBHs merge with isotropic spin tilts, we expect 16\% of 1g+2g or 2g+2g HMs to have $\chi_\mathrm{eff} < -0.3$. Therefore, {\bf HMs make up no more than 26\% of the BBH population (90\% credibility).}
    \item In agreement with theoretical expectations, the data permit the HM fraction to increase with BBH primary mass. However, this is not required by the data. {\bf HMs may dominate the BBH population with primary masses above $\sim60\,M_\odot$, but such massive BBHs make up less than 1\% of the population} (see Fig.~\ref{fig:negchieff-linevolm1}). This implies that if the HM fraction is as high as 26\%, the vast majority of these HMs have $m_1 < 60\,M_\odot$.
    \item Below primary masses of $\sim60\,M_\odot$, HMs make up the minority of systems. {\bf This suggests that stellar collapse can create BHs up to $\gtrsim60\,M_\odot$}, implying a lower limit on the pair-instability mass gap. Alternatively, more exotic possibilities may populate the pair-instability gap, or our assumption of an isotropic spin tilt distribution may be invalid. However, if HMs between preferentially aligned BBHs populate the gap, the dearth of systems with $\chi_\mathrm{eff} > 0.4$ limits their contribution.
    \item If HMs contribute significantly to the BBH population, {\bf we expect to observe a handful of systems with $\chi_\mathrm{eff} < -0.3$ within the LVK's next observing run (see Fig.~\ref{fig:chineg_obs_vtotal_HMfrac.pdf}). We will probably be able to identify these events as confidently ($> 99\%$ credibility) having negative $\chi_\mathrm{eff}$.} If the HM fraction is above 10\%, we expect to observe a BBH with $\chi_\mathrm{eff} < -0.3$ within the first 150 events.
\end{itemize}

\acknowledgments
We thank Sylvia Biscoveanu, Christopher Berry, and Reed Essick for their helpful comments on the manuscript. MF is supported by NASA through NASA Hubble Fellowship grant HST-HF2-51455.001-A awarded by the Space Telescope Science Institute, which is operated by the Association of Universities for Research in Astronomy, Incorporated, under NASA contract NAS5-26555. 
CK is supported by the Riedel Family Fellowship.
VK is grateful for support from a Guggenheim Fellowship, from CIFAR as a Senior Fellow, and from Northwestern University, including the Daniel I. Linzer Distinguished University Professorship fund. 
This material is based upon work supported by NSF's LIGO Laboratory which is a major facility fully funded by the National Science Foundation.

\bibliographystyle{aasjournal}
\bibliography{references}
\end{document}